\documentclass[10pt,conference,twocolumn,letterpaper]{IEEEtran}
\IEEEoverridecommandlockouts

\usepackage{amsmath}
\usepackage{amsfonts}
\usepackage{amssymb}
\usepackage{amsthm}
\usepackage{tabularx}
\usepackage{mathrsfs} 
\usepackage{soul}
\usepackage{graphicx}

\usepackage{url}
\usepackage{hyperref}

\usepackage{stfloats}
\usepackage{float}
\usepackage{graphicx}
\usepackage{cite}
\usepackage{xcolor}
\usepackage{subfigure}

\usepackage{cleveref}

\usepackage{threeparttable}

\usepackage{algorithm}
\usepackage{algpseudocode}

\usepackage{pifont}

\usepackage{booktabs}

\usepackage{multirow}

\usepackage{rotating}

\usepackage{makecell}

\usepackage{array}

\makeatletter
\def\ps@headings{}                
\makeatother

\def\BibTeX{{\rm B\kern-.05em{\sc i\kern-.025em b}\kern-.08em
    T\kern-.1667em\lower.7ex\hbox{E}\kern-.125emX}}
\usepackage{balance}

\begin{document}
\title{
Ambient Backscatter Communication Assisted by Fluid Reconfigurable Intelligent Surfaces}
\author{
\IEEEauthorblockN{
Masoud Kaveh\IEEEauthorrefmark{2},
Farshad Rostami Ghadi\IEEEauthorrefmark{4},
Riku J\"antti\IEEEauthorrefmark{2},
Kai-Kit Wong\IEEEauthorrefmark{3},
F. Javier Lopez-Martinez\IEEEauthorrefmark{4}
}
\IEEEauthorblockA{\IEEEauthorrefmark{2}Department of Information and Communications Engineering, Aalto University, Espoo, Finland}
\IEEEauthorblockA{\IEEEauthorrefmark{4}Department of Signal Theory, Networking and Communications, University of Granada, Granada, Spain. }
\IEEEauthorblockA{\IEEEauthorrefmark{3}Department of Electronic and Electrical Engineering, University College London, London, U.K.}

}


\maketitle

\begin{abstract}
This paper investigates the integration of a fluid reconfigurable intelligent surface (FRIS) into ambient backscatter communication (AmBC) systems. Unlike conventional reconfigurable intelligent surfaces (RISs) with fixed position elements, FRIS employs fluidic elements that can dynamically adjust their positions, offering enhanced spatial adaptability. We develop a system model where an AmBC tag communicates with a reader through an FRIS, which is particularly beneficial in scenarios where the direct tag-to-reader link is weak or blocked by obstacles. The achievable backscatter rate is analyzed, and the optimization of FRIS element positions is formulated as a non-convex problem. To address this, we employ particle swarm optimization (PSO) to obtain near-optimal configurations of the fluid elements. Simulation results demonstrate that FRIS-aided AmBC significantly outperforms conventional RIS-based AmBC systems in terms of achievable throughput. 
\end{abstract}
\vspace{6pt}

\section{Introduction}
\IEEEPARstart{A}{mbient} backscatter communication (AmBC) has emerged as a key enabler for ultra-low-power Internet of Things (IoT), allowing tags to reflect ambient signals rather than generating their own \cite{Duan_ComMag2020}. This paradigm drastically reduces energy consumption and hardware complexity, making it attractive for dense, battery-free sensing at scale \cite{Kaveh2025:VPD-PLA}. However, AmBC performance is often limited by severe double-fading on the cascaded paths, weak or blocked direct links in cluttered indoor settings, and narrow link budgets that yield low singnal-to-noise ratio (SNR) and modest data rates. These constraints motivate environment-aware solutions that can shape propagation to stabilize links and lift achievable rate without sacrificing the energy neutrality of the tag \cite{Basharat_WCM2022}.

Reconfigurable intelligent surfaces (RISs) were introduced precisely to reprogram the wireless channel and have since been explored to elevate AmBC performance \cite{Kaveh2024:SecureRIS}. Early work combined RIS with AmBC so that phase compensation at the surface mitigates multipath distortion and improves throughput \cite{Nemati_WCNC2020}. RIS-assisted AmBC has been extended to space–air–ground integrated IoT, where lightweight phase control strategies boost achievable rate \cite{Liu_IoTJ2022}. To further raise spectral efficiency under tight power budgets, RIS-aided AmBC with advanced spatial modulation was proposed, showing notable SNR gains over conventional schemes \cite{Bhowal_TGCN2021}. The RIS concept also pairs well with multiple-access in AmBC \cite{Usman_ICDCSW2022,Zuo_TVT2021}. Furthermore, RIS-aided AmBC has been used to co-optimize energy harvesting, beamforming, and power allocation for throughput gains in practical energy harvesting models \cite{Ramezani_TVT2021}. 

Despite these gains, RIS-aided AmBC systems face key hurdles: multiplicative path loss from the cascaded link, rapidly growing channel state information (CSI) overhead with more elements, and the geometric rigidity of fixed planar lattices that limits spatial-diversity gains \cite{elmossallamy2020ris}. 
Building on these limitations, a natural next step is to bring fluidity to the surface itself. 
The inspiration for this direction can be traced back to the fluid antenna system (FAS), where terminals dynamically reconfigure their active ports to exploit spatial diversity within a compact aperture \cite{Wong2021:FAS, Fluid-survey}. 
FAS and RIS have also coexisted in the same system setup, leveraging both fluid reconfigurability at the user side and intelligent reflection at the surface \cite{Ghadi2024:RIS_FAS}. 
Extending this paradigm from terminals to metasurfaces, the notion of bringing fluidity into the reflecting aperture itself has only recently been formalized under the framework of fluid RIS (FRIS), where the reflecting units behave as fluidic subelements that can reposition within a predefined aperture, adding geometry control on top of phase control \cite{Salem2025:FRIS_FirstLook,Xiao2025:FRIS_Joint}. This concept has further evolved into FIRES, which combines reflecting and emitting modes to approach 360$^\circ$ coverage \cite{Ghadi2025:FIRES}. Analytical studies also report closed-form ergodic-capacity and outage characterizations for FRIS-aided links, showing improvements over RISs \cite{Ghadi2025:FRIS_Arxiv}. 

These features make FRIS a natural fit for AmBC, as it preserves the passive and low-power operation of RIS while introducing the positional agility needed to overcome double-fading and improve throughput in AmBC setups. Despite this potential, FRIS has remained unexplored in the AmBC context. Motivated by this gap, we propose an FRIS-aided AmBC system in which a passive tag harvests energy from an ambient source and communicates with a reader through the assistance of FRIS. On top of this model, we formulate the achievable rate maximization problem, where the key design variable is the positioning of fluid reflective elements across the metasurface. To address the non-convexity of this problem, we employ particle swarm optimization (PSO) to efficiently search over possible spatial arrangements, enabling FRIS to exploit its positional flexibility for AmBC link strengthening. Our results demonstrate that FRIS-aided AmBC outperforms conventional RIS-based AmBC in term of achievable rate.

\section{System and Channel Models} \label{sec_sys}
\begin{figure}[t]
    \centering    \includegraphics[width=0.45\textwidth]{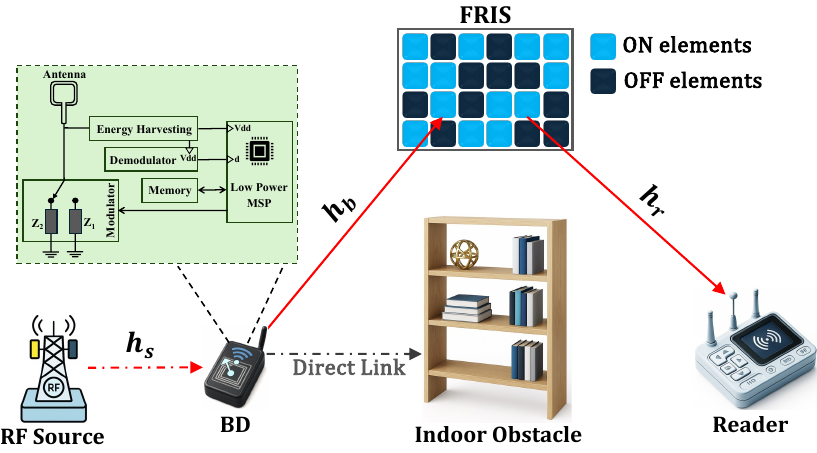}
    \caption{The considered FRIS-aided AmBC system with the FRIS comprising ON-OFF unit elements.}
    \label{fig:sysmodel}
\end{figure}

As illustrated in Fig. \ref{fig:sysmodel}, we consider an AmBC system link where a radio frequency (RF) source illuminates a passive backscatter device (BD) or a tag that modulates and reflects information toward a reader through an FRIS. We assume that the direct link between the tag and the reader is weak or blocked by an indoor obstacle unless otherwise indicated. The FRIS occupies an aperture $\mathcal{A}=[0,W_\mathrm{x}\lambda]\times[0,W_\mathrm{z}\lambda]$ (in meters) and contains $M=M_\mathrm{x}M_\mathrm{z}$ fluid meta-elements arranged on a dense grid of candidate positions. At a given coherence interval, only $M_\mathrm{o}$ elements are activated (ON) while the rest are OFF (matched load). 
We denote the element coordinates by
$
\mathbf{p}_m\triangleq(x_m,z_m)\in\mathcal{A}, \ m=1,\ldots,M,
$
and collect them in \(\mathbf{P}\!=\![\mathbf{p}_1^\top,\ldots,\mathbf{p}_M^\top]^\top\). 
The ON subset is captured by a \emph{selection matrix} \(\mathbf{S}_{M_{\mathrm{o}}}\in\{0,1\}^{M\times M_{\mathrm{o}}}\) that extracts the active entries (i.e., columns of \(\mathbf{I}_M\)); hence \(\mathbf{S}_{M_{\mathrm{o}}}^{\!\top}\mathbf{S}_{M_{\mathrm{o}}}=\mathbf{I}_{M_{\mathrm{o}}}\) and \(\mathbf{S}_{M_{\mathrm{o}}}\mathbf{S}_{M_{\mathrm{o}}}^{\!\top}\) is a diagonal ON/OFF mask.

Denote by $h_\mathrm{s}\!\in\!\mathbb{C}$ the channel coefficient between the RF source and the tag, by $\mathbf{h}_\mathrm{b}\!\in\!\mathbb{C}^{M\times 1}$ the channel coefficient between the tag and the FRIS (one entry per FRIS element) and by $\mathbf{h}_\mathrm{r}\!\in\!\mathbb{C}^{M\times 1}$ the channel coefficient between the FRIS and the reader. All links follow large–scale path loss $L(d)=\rho d^{-\alpha}$ and small–scale Rician fading
\begin{equation}
\mathbf{q}
=\sqrt{L}\!\left(\sqrt{\tfrac{K}{K+1}}\;\mathbf{q}_{\mathrm{LoS}}
+\sqrt{\tfrac{1}{K+1}}\;\mathbf{J}^{1/2}\mathbf{w}\right),\  
\mathbf{w}\sim\mathcal{CN}(\mathbf{0},\mathbf{I}_M),
\label{eq:rician}
\end{equation}
for $\mathbf{q}\in\{\mathbf{h}_\mathrm{b},\mathbf{h}_\mathrm{r}\}$ (and the scalar form for $h_\mathrm{s}$). The LoS steering vector at the FRIS follows the element coordinates and hop angles; e.g., its \(m\)-th entry is
\begin{equation}
[\mathbf{q}_{\mathrm{LoS}}]_m
=\exp\!\Big(j\tfrac{2\pi}{\lambda}\big(x_m\sin\psi_{\mathrm{az}}\cos\psi_{\mathrm{el}}+z_m\sin\psi_{\mathrm{el}}\big)\Big).
\end{equation}
and $\mathbf{q}_{\text{LoS}}$ is a steering vector built from the departure/arrival angles of the corresponding hop. To capture inter-element coupling on a dense aperture, we adopt a Jakes' spatial correlation
matrix \(\mathbf{J}\in\mathbb{C}^{M\times M}\) with entries
\begin{align}
[\mathbf{J}]_{i,j}
&= J_0\!\Big(\tfrac{2\pi}{\lambda}\,d_{i,j}\Big), \label{eq:J-bessel} \\
d_{i,j}^2
&= d_{\mathrm{x}}^2\big(i_{\mathrm{x}}-j_{\mathrm{x}}\big)^2
   + d_{\mathrm{z}}^2\big(i_{\mathrm{z}}-j_{\mathrm{z}}\big)^2, \nonumber
\end{align}
where \(J_0(\cdot)\) is the zeroth-order Bessel function, \(d_{\mathrm{x}}=\tfrac{W_{\mathrm{x}}\lambda}{M_{\mathrm{x}}}\), \(d_{\mathrm{z}}=\tfrac{W_{\mathrm{z}}\lambda}{M_{\mathrm{z}}}\), and the index map
\(i_{\mathrm{x}}=\bmod(i\!-\!1,M_{\mathrm{x}})\), \(i_{\mathrm{z}}=\big\lfloor\tfrac{i-1}{M_{\mathrm{x}}}\big\rfloor\) (similarly for \(j\)).  
We use \(\mathbf{J}^{1/2}\) (e.g., Cholesky factor) in \eqref{eq:rician} to color the i.i.d. vector \(\mathbf{w}\).
This construction is consistent with recent FRIS analyses and enables tractable modeling of fluid repositioning through $\{(x_m,z_m)\}$ (see, e.g., FRIS performance modeling and FRIS/FIRES geometry and correlation discussions in \cite{Ghadi2025:FIRES,Ghadi2025:FRIS_Arxiv}).

The BD employs binary backscatter via impedance switching with effective reflection coefficient difference $\Delta\Gamma$; we absorb its magnitude into $\alpha\!\in\!(0,1]$ and its symbol into $\beta\in\{\pm1\}$. The FRIS applies a diagonal reflection matrix
\begin{equation}
\boldsymbol{\Phi}=\mathrm{diag}\!\big(a_1e^{j\phi_1},\dots,a_Me^{j\phi_M}\big),
\end{equation}
where $a_\text{m}\in\{0,1\}$ selects ON/OFF and $\phi_\text{m}$ are per-element phases. For any given fluid layout and ON set, coherent combining is achieved by the optimal choice
$\phi_\text{m}^\star=\angle h_\text{b,m}-\angle h_\text{r,m}$.


Let the source transmit a unit-power symbol $x$ with power $P_\text{s}$. Neglecting the direct (BD$\to$reader) path, the received baseband signal at the reader is
\begin{equation}
y_{\mathrm{r}}
=\sqrt{P_\text{s}\,L_{\mathrm{s}}L_{\mathrm{b}}L_{\mathrm{r}}}\;\alpha\,h_{\mathrm{s}}\,
\underbrace{\mathbf{h}_{\mathrm{r}}^{\sf H}\mathbf{J}^{\frac{1}{2}}
\mathbf{S}_{M_{\mathrm{o}}}^{\!\top}\boldsymbol{\Phi}\,
\mathbf{S}_{M_{\mathrm{o}}}\mathbf{J}^{\frac{1}{2}}\mathbf{h}_{\mathrm{b}}}_{\displaystyle H_{\mathrm{eq}}}
\;\beta\,s + z,
\label{eq:rx}
\end{equation}
with the additional white Gaussian noise (AWGN) $z\sim\mathcal{CN}(0,\sigma^2)$. Including path losses of the cascaded hops, and defining the average transmit SNR as \(\bar{\gamma}\triangleq P_\text{s}/\sigma^2\), the instantaneous SNR is
\begin{equation}
\gamma_{\mathrm{r}}
= \bar{\gamma}\,\alpha^2\,L_{\mathrm{s}}L_{\mathrm{b}}L_{\mathrm{r}}\,|h_{\mathrm{s}}|^2\,|H_{\mathrm{eq}}|^2.
\label{eq:SNR}
\end{equation}
Accordingly, the achievable backscatter rate (Shannon bound) is
\begin{equation}
R_\text{r}=\log_2\!\big(1+\gamma_\text{r}\big).
\end{equation}
Under the optimal phase choice \(\phi_m^\star=\angle h_{\mathrm{b},m}-\angle h_{\mathrm{r},m}\) on the \emph{selected} entries, \(|H_{\mathrm{eq}}|\) reduces to the coherent sum magnitude
\(|H_{\mathrm{eq}}|=\sum_{m\in\mathcal{O}}|h_{\mathrm{r},m}|\,|h_{\mathrm{b},m}|\),
where \(\mathcal{O}\) indexes the \(M_{\mathrm{o}}\) ON elements picked by \(\mathbf{S}_{M_{\mathrm{o}}}\).

The FRIS’s \textit{fluidity} enters through the element coordinates
$\mathbf{P}=\{\mathbf{p}_m\}_{m=1}^{M}$ with $\mathbf{p}_m=(x_m,z_m)\in\mathcal{A}$,
and through the ON subset represented by the selection matrix
$\mathbf{S}\in\mathcal{S}_{M_{\mathrm{o}}}$ (whose columns are distinct canonical
vectors so that $\mathbf{S}^{\!\top}\mathbf{S}=\mathbf{I}_{M_{\mathrm{o}}}$).
We assume a fixed ON budget $M_{\mathrm{o}}$ and search over positions within
$\mathcal{A}$ (or over a dense preset grid) subject to a minimum inter-element
spacing $d_{\min}$ to mitigate coupling,
\begin{equation}
\|\mathbf{p}_m-\mathbf{p}_n\|_2 \ge d_{\min},\qquad \forall\, m\neq n .
\label{eq:minspacing}
\end{equation}
Let $\mathcal{F}$ denote the feasible set of pairs $(\mathbf{P},\mathbf{S})$
satisfying the aperture, spacing, and cardinality constraints above. The design
problem used later is to \emph{maximize} the rate
$R_{\mathrm{r}}(\mathbf{P},\mathbf{S})$ over
$(\mathbf{P},\mathbf{S})\in\mathcal{F}$. 
It is worth noting that since the objective of this paper is to establish a fair benchmark and isolate the \emph{positional} gains of FRIS over a conventional RIS, we do not optimize phases as free variables; instead, the reflective elements are assigned their closed-form, rate-maximizing alignment, i.e.,
$\boldsymbol{\Phi}=\mathrm{diag}\!\big(e^{j\phi_1^\star},\ldots,e^{j\phi_{M_{\mathrm{o}}}^\star}\big),
\ \phi_m^\star=\angle h_{\mathrm{b},m}-\angle h_{\mathrm{r},m}.$
Accordingly, only the fluid-element locations \(\mathbf{P}\) and the activation pattern (the selection matrix \(\mathbf{S}\)) are optimized.

\section{Problem Formulation}\label{sec:problem}

In this section, we focus on maximizing the achievable backscattering rate of the proposed FRIS-enabled AmBC system. 
Let the element coordinates be
\(
\mathbf{P}\triangleq[\mathbf{p}_1^\top,\ldots,\mathbf{p}_M^\top]^\top
\in\mathbb{R}^{2M},\ \mathbf{p}_m=(x_m,z_m)\in\mathcal{A},
\)
and let the ON subset be represented by a selection matrix
\(
\mathbf{S}\in\mathcal{S}_{M_{\mathrm{o}}}
\triangleq\{\mathbf{S}\in\{0,1\}^{M\times M_{\mathrm{o}}}:
\mathbf{S}^{\!\top}\mathbf{S}=\mathbf{I}_{M_{\mathrm{o}}}\}.
\)
Aperture and spacing constraints define the geometric feasible set
\begin{equation}
\mathcal{P}\!=\!\Big\{\mathbf{P}:\ \mathbf{p}_m\in\mathcal{A},\
\|\mathbf{p}_m-\mathbf{p}_n\|_2\ge d_{\min},\ \forall m\neq n\Big\}.
\label{eq:setP}
\end{equation}
The RF source power \(P_{\mathrm{s}}\) is treated as a parameter obeying the regulation \(0<P_{\mathrm{s}}\le P_{\max}\).
With \(\bar{\gamma}\triangleq P_{\mathrm{s}}/\sigma^2\), the instantaneous reader rate under optimal phasing is
\begin{equation}
R_{\mathrm{r}}(\mathbf{P},\mathbf{S})
=\log_2\!\Big(1+ \bar{\gamma}\,\alpha^2 L_{\mathrm{s}}L_{\mathrm{b}}L_{\mathrm{r}}
\,|h_{\mathrm{s}}|^2 \,|H_{\mathrm{eq}}(\mathbf{P},\mathbf{S})|^2\Big),
\label{eq:instRatePF}
\end{equation}
where \(H_{\mathrm{eq}}(\mathbf{P},\mathbf{S})=
\mathbf{h}_{\mathrm{r}}^{\sf H}\mathbf{J}^{\frac{1}{2}}\mathbf{S}^{\!\top}\boldsymbol{\Phi}^\star
\mathbf{S}\mathbf{J}^{\frac{1}{2}}\mathbf{h}_{\mathrm{b}}\) as in \eqref{eq:rx}, and the dependence on \(\mathbf{P}\) propagates through \(\mathbf{J}(\mathbf{P})\), \(\mathbf{h}_{\mathrm{b}}(\mathbf{P})\), and \(\mathbf{h}_{\mathrm{r}}(\mathbf{P})\).
For a fixed \(P_{\mathrm{s}}\), the design problem is
\begin{subequations}\label{eq:P1}
\begin{align}
\text{(P1)}\quad
\max_{\mathbf{P},\,\mathbf{S}}\ \ & R_{\mathrm{r}}(\mathbf{P},\mathbf{S}) \label{eq:P1a}\\
\text{s.t.}\quad &
\mathbf{P}\in\mathcal{P}, \label{eq:P1b}\\
&\mathbf{S}\in\mathcal{S}_{M_{\mathrm{o}}}, \label{eq:P1c}\\
& \|\mathbf{p}_m-\mathbf{p}_n\|_2\ge d_{\min},\ \forall\, m\neq n. \label{eq:P1d}
\end{align}
\end{subequations}
To account for small–scale fading, we maximize the sample–average rate
\begin{subequations}\label{eq:P2}
\begin{align}
\text{(P2)}\quad
\max_{\mathbf{P},\,\mathbf{S}}\ \ &
\frac{1}{N}\sum_{n=1}^{N} R_{\mathrm{r}}^{(n)}(\mathbf{P},\mathbf{S}) \label{eq:P2a}\\
\text{s.t.}\quad &
\mathbf{P}\in\mathcal{P}, \label{eq:P2b}\\
&\mathbf{S}\in\mathcal{S}_{M_{\mathrm{o}}}, \label{eq:P2c}\\
& \|\mathbf{p}_m-\mathbf{p}_n\|_2\ge d_{\min},\ \forall\, m\neq n. \label{eq:P2d}
\end{align}
\end{subequations}
where $N$ is the number of i.i.d. channel draws used to obtain the sample-average approximation (SAA) and \(R_{\mathrm{r}}^{(n)}\) is evaluated from i.i.d. channel draws shaped by the Jakes' correlation induced by \(\mathbf{P}\).
Problems \eqref{eq:P1}–\eqref{eq:P2} are non-convex due to the discrete selection \(\mathbf{S}\), geometry–dependent correlation \(\mathbf{J}(\mathbf{P})\), and the cascaded channel. We therefore adopt PSO algorithm to search over the fluid degrees of freedom.

\paragraph*{Encoding}  
We use either (i) a \emph{general activation} encoding in which a particle state is
\(
\mathbf{x}=[\mathbf{P}^\top,\ \boldsymbol{\nu}^\top]^\top
\)
with \(\boldsymbol{\nu}\in\mathbb{R}^M\) relaxed and mapped to a selector \(\mathbf{S}(\boldsymbol{\nu})\) by picking the \(M_{\mathrm{o}}\) largest entries; or (ii) a \emph{contiguous-mask} encoding where the ON set is a rectangular mask of size \(m_{\mathrm{x}}\!\times m_{\mathrm{z}}\) (\(m_{\mathrm{x}}m_{\mathrm{z}}=M_{\mathrm{o}}\)) anchored at \(\boldsymbol{\kappa}=(i_0,j_0)\) on a dense grid, so the particle state is \(\mathbf{x}=[\mathbf{P}^\top,\ \boldsymbol{\kappa}^\top]^\top\).
Both variants keep phases fixed to \(\boldsymbol{\Phi}^\star\).

\paragraph*{Fitness}  
For (P1) we evaluate the instantaneous fitness \(f(\mathbf{x})=R_{\mathrm{r}}(\mathbf{P},\mathbf{S})\) via \eqref{eq:instRatePF}; for (P2) we use the SAA fitness \(\frac{1}{N}\sum_{n}R_{\mathrm{r}}^{(n)}(\mathbf{P},\mathbf{S})\).  
To handle geometry, we optionally employ a penalized fitness
\begin{equation}
\tilde f(\mathbf{x})
= f(\mathbf{x})
-\tau\Big(\mathcal{B}_{\mathrm{space}}(\mathbf{P})
+\mathcal{B}_{\mathrm{apert}}(\mathbf{P})\Big),
\label{eq:penalty}
\end{equation}
with a large \(\tau>0\); 
\(
\mathcal{B}_{\mathrm{space}}
=\sum_{m<n}\max\{0,\,d_{\min}-\|\mathbf{p}_m-\mathbf{p}_n\|_2\}^{2}
\)
and
\(
\mathcal{B}_{\mathrm{apert}}
=\sum_{m}\mathrm{dist}^{2}(\mathbf{p}_m,\mathcal{A})
\),
where \(\mathrm{dist}(\cdot,\mathcal{A})=0\) for in-bounds points.
(For the contiguous-mask variant, feasibility is enforced directly by bounding \(\boldsymbol{\kappa}\), so penalties are unnecessary.)

\paragraph*{Iterations and projection}  
Let \(\mathbf{x}_i^{(t)}\) and \(\mathbf{v}_i^{(t)}\) be the position and velocity of particle \(i\) at iteration \(t\). The updates follow 
\begin{align}
\mathbf{v}_i^{(t+1)}& \hspace{-2pt}=\hspace{-2pt}
\omega \mathbf{v}_i^{(t)}
\hspace{-2pt}+\hspace{-2pt}c_1\mathbf{r}_1 \hspace{-2pt} \odot \hspace{-2pt} \big(\mathbf{x}_{i,\mathrm{best}} \hspace{-2pt}-\hspace{-2pt} \mathbf{x}_i^{(t)}\big)
\hspace{-2pt} +\hspace{-2pt} c_2\mathbf{r}_2 \hspace{-2pt} \odot \hspace{-2pt} \big(\mathbf{x}_{\mathrm{best}} \hspace{-2pt}- \hspace{-2pt} \mathbf{x}_i^{(t)}\big), \\
\mathbf{x}_i^{(t+1)}&=\Pi_{\mathcal{X}}\!\left(\mathbf{x}_i^{(t)}+\mathbf{v}_i^{(t+1)}\right),
\end{align}
where \(\omega\) is the inertia weight, \(c_1,c_2\) are cognitive/social gains, \(\mathbf{r}_1,\mathbf{r}_2\sim\mathcal{U}(0,1)\) (elementwise), and \(\Pi_{\mathcal{X}}(\cdot)\) projects onto the constraint set \(\mathcal{X}\) (i.e., \(\mathcal{P}\times\mathcal{S}_{M_{\mathrm{o}}}\) for the general case or \(\mathcal{P}\times\mathcal{K}\) for the contiguous-mask anchors). Personal and global bests are updated using \(f(\cdot)\) or \(\tilde f(\cdot)\). Algorithm~\ref{alg:pso-fris-short} summarizes the proposed PSO solver for (P1)/(P2).


\begin{algorithm}[t]
\caption{PSO for FRIS-aided AmBC (Optimize Positions $\mathbf{P}$ and Activation $\mathbf{S}$)}
\label{alg:pso-fris-short}
\small
\begin{algorithmic}[1]
\Require Aperture $\mathcal{A}$, spacing $d_{\min}$, ON budget $M_{\mathrm{o}}$; draws $N$; swarm $N_p$; iters $T$; PSO $(\omega,c_1,c_2)$; system params $(P_{\mathrm{s}},\alpha,\sigma^2,L_{\mathrm{s}},L_{\mathrm{b}},L_{\mathrm{r}})$; Rician \& $\mathbf{J}(\cdot)$ model; \textsc{Mode}$\in\{$general, mask$\}$.
\Ensure Best $(\mathbf{P}^\star,\mathbf{S}^\star)$ and $R_{\mathrm r}^\star$.
\State \textbf{Init} each particle $p$: sample feasible $\mathbf{P}_p^{(0)}\!\in\!\mathcal P$; set activation state (general: scores $\boldsymbol{\nu}_p^{(0)}$; mask: anchor $\boldsymbol{\kappa}_p^{(0)}$); velocities $\!=\!0$; pbest $\leftarrow$ init; global best $\leftarrow$ particle 1.
\For{$t=0$ to $T-1$} \label{line:outer}
  \For{each particle $p$}
    \State \textbf{Decode} activation: 
    $\mathbf{S}_p^{(t)}\!\leftarrow\!\begin{cases}
      \text{top-$M_{\mathrm{o}}$ of }\boldsymbol{\nu}_p^{(t)} & \text{(general)}\\
      \text{mask from }\boldsymbol{\kappa}_p^{(t)} & \text{(mask)}
    \end{cases}$
    \State Build $\mathbf{J}_p=\mathbf{J}(\mathbf{P}_p^{(t)})$.
    \State \textbf{SAA fitness:} For $n=1\!\dots\!N$, draw $h_{\mathrm s}^{(n)},\mathbf{h}_{\mathrm b}^{(n)},\mathbf{h}_{\mathrm r}^{(n)}$ (Rician with $\mathbf{J}_p$); set optimal phases on selected entries $\phi_m^\star=\angle h_{\mathrm b,m}^{(n)}-\angle h_{\mathrm r,m}^{(n)}$; compute $R^{(n)}$ via \eqref{eq:SNR}. Set $f_p=\frac{1}{N}\sum_n R^{(n)}$.
    \State (Optional) $\tilde f_p=f_p-\tau\,[\mathcal B_{\text{space}}(\mathbf{P}_p^{(t)})+\mathcal B_{\text{apert}}(\mathbf{P}_p^{(t)})]$.
    \State \textbf{Update} pbest / gbest with $f_p$ (or $\tilde f_p$).
  \EndFor
  \State \textbf{PSO step:} For each $p$, 
  $\mathbf{v}_p^{(t+1)}\!=\!\omega\mathbf{v}_p^{(t)}\!+\!c_1\mathbf{r}_1\!\odot(\mathbf{x}_{p,\mathrm{best}}\!-\!\mathbf{x}_p^{(t)})\!+\!c_2\mathbf{r}_2\!\odot(\mathbf{x}_{\mathrm{best}}\!-\!\mathbf{x}_p^{(t)})$; 
  $\mathbf{x}_p^{(t+1)}\!=\!\Pi_{\mathcal X}(\mathbf{x}_p^{(t)}\!+\!\mathbf{v}_p^{(t+1)})$,\\
  where $\mathbf{x}_p=[\mathbf{P}_p^\top,\boldsymbol{\nu}_p^\top]^\top$ (general) or $[\mathbf{P}_p^\top,\boldsymbol{\kappa}_p^\top]^\top$ (mask), $\Pi_{\mathcal X}$ projects onto aperture \& spacing (and, for mask, anchor bounds).
\EndFor
\State \textbf{Return} $(\mathbf{P}^\star,\mathbf{S}^\star)$ and $R_{\mathrm r}^\star$.
\end{algorithmic}
\end{algorithm}

\section{ Results and Discussions} \label{sec_sim}

To assess the performance of the proposed FRIS-aided AmBC system, we conduct simulations under independent channel realizations for each transmit power point. The carrier frequency is set to $f_c = 3.5$~GHz, corresponding to a wavelength $\lambda = 0.0857$~m. The FRIS aperture spans $3\lambda \times 3\lambda$ in both horizontal and vertical dimensions, discretized into a dense grid of $M_x = M_z = 20$ fluid elements, yielding a total of $M = 400$ candidate positions. Out of these, only $M_{\mathrm{o}} \in \{25, 100, 225\}$ elements are adaptively activated using the proposed PSO-based selection. For a fair comparison, the conventional RIS baseline employs a fixed $m \times m$ array with $\hat{M} = M_\mathrm{o}$ elements and uniform $\lambda/2$ spacing. The path-loss exponent is set to $\alpha = 2.5$, and the large-scale attenuation between the source, tag (BD), FRIS/RIS, and reader follows $L(d) = d^{-\alpha}$. Small-scale fading is modeled as correlated Rician with $K=5$, where the correlation across the surface is generated via a Bessel $J_0$ kernel and enforced by Cholesky factorization. BD employs binary reflection with amplitude $\alpha=1$, while the direct BD-to-reader path is excluded unless otherwise stated. The additive noise variance is $\sigma^2 = 10^{-9}$. The maximum source power is set to $30$~dB, and achievable backscatter rates are averaged across all runs. For the FRIS optimization, the PSO controller uses $N_p = 50$ particles and $50$ iterations, with inertia weight $\omega=0.6$ and cognitive/social coefficients $c_1=c_2=1.2$. This setup allows us to isolate the array-gain improvements of FRIS positioning against a conventional RIS baseline under optimal phase alignment.

\begin{figure}[t]
    \centering    \includegraphics[width=0.37\textwidth]{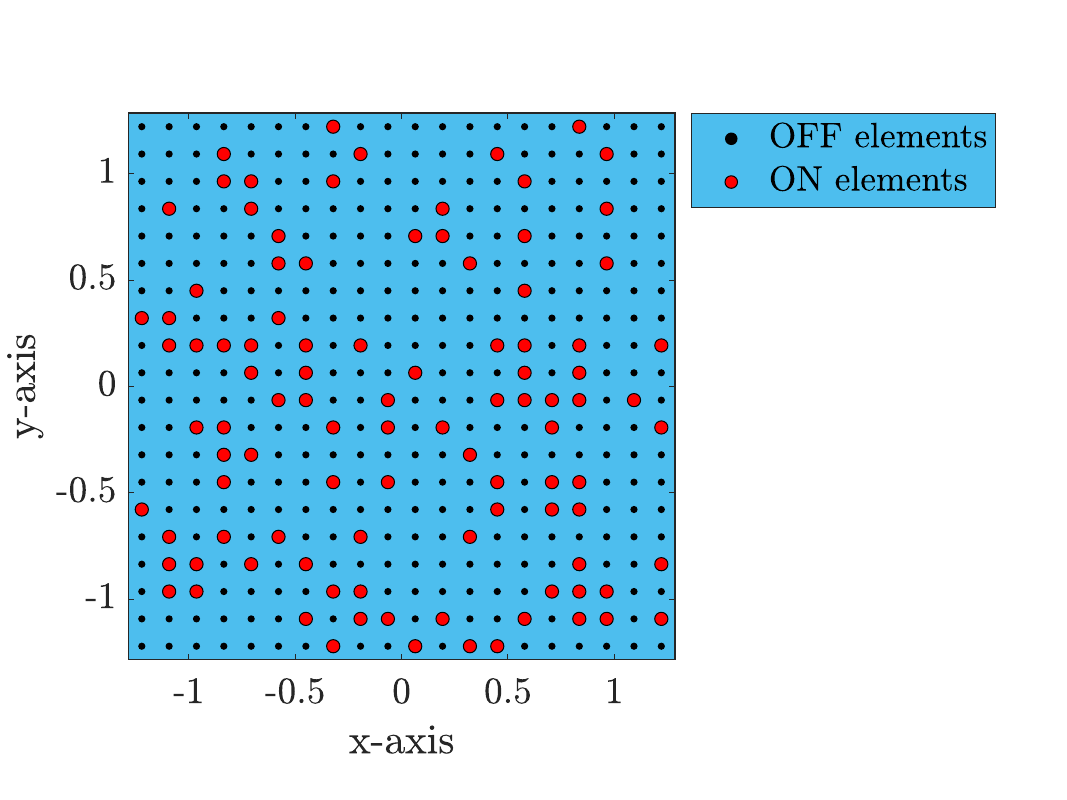}
    \caption{The system configuration for $(M_\mathrm{x},M_\mathrm{z})=(20,20)$ and $M_\mathrm{o}=100$.}
    \label{fig:axis}
\end{figure}

Fig. \ref{fig:axis} shows the ON elements selected by the PSO algorithm on a $20 \times 20$ FRIS grid with $M_{\mathrm{o}}=100$. The ON elements (red) are distributed across the aperture, while the OFF elements (black) remain inactive. This visualization mainly illustrates the flexibility of FRIS in activating arbitrary subsets of elements, as opposed to the fixed structure of conventional RIS.

\begin{figure}[t]
    \centering    \includegraphics[width=0.37\textwidth]{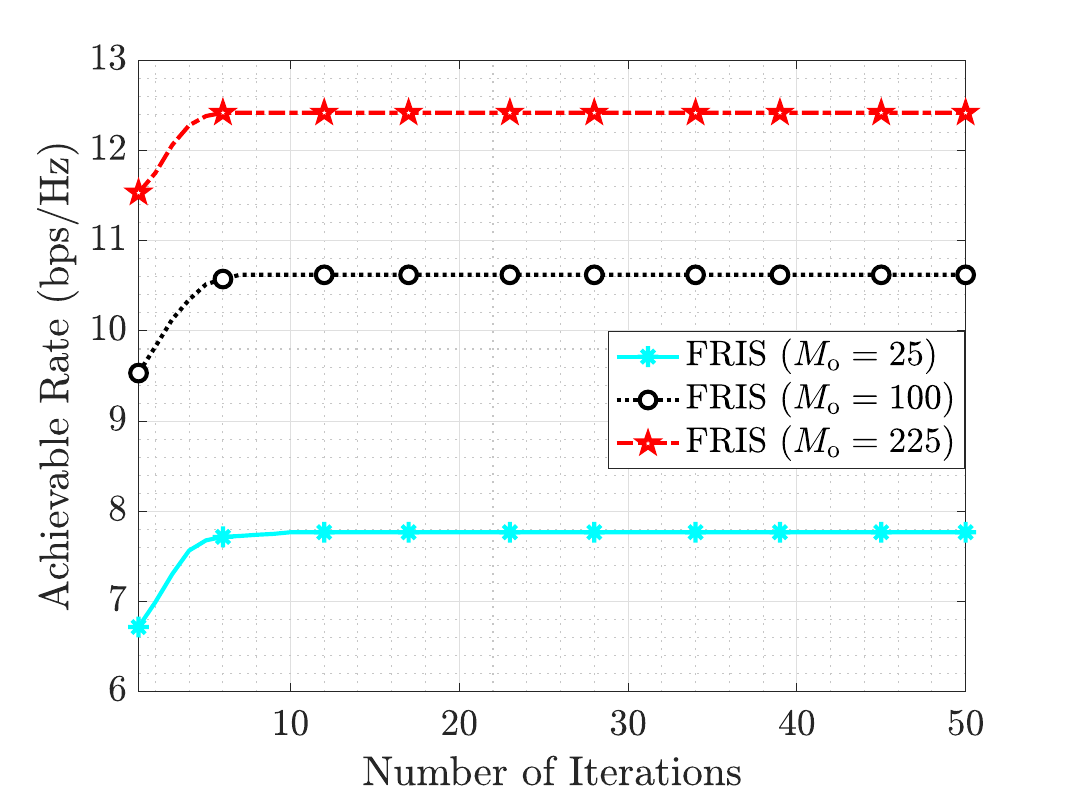}
    \caption{PSO convergence of the FRIS achievable rate for different $M_\mathrm{o}$ and $\bar{\gamma}_\mathrm{r}=10$ dB.}
    \label{fig:iter}
\end{figure}
Fig.~\ref{fig:iter} plots the FRIS achievable rate versus the PSO iteration index for three ON budgets $M_{\mathrm{o}}\!\in\!\{25,100,225\}$ at $\bar\gamma_{\mathrm{r}}=10$\,dB. In all cases, the curves exhibit a steep ascent during the first few iterations and then quickly flatten, converging within $\approx 8$--$12$ iterations; beyond that point the gain per iteration is negligible ($<0.05$\,bps/Hz). This behavior comes from the rapid relocation of the contiguous $m\!\times\! m$ active block toward the dominant spatial cluster induced by the Jakes' correlation, while the global–best update guarantees a monotone, non-decreasing best-so-far rate. As expected, larger $M_{\mathrm{o}}$ yields higher steady-state rates due to stronger coherent combining; the magnitude $\big|\sum_{m\in\mathcal{O}} |h_{\mathrm{r},m}||h_{\mathrm{b},m}|\big|$ grows with $M_{\mathrm{o}}$, yet the incremental benefit diminishes (\emph{e.g.}, final rates are about $7.8$, $10.6$, and $12.4$\,bps/Hz for $M_{\mathrm{o}}=25,100,225$, respectively), consistent with the logarithmic rate law and spatial correlation. 
Furthermore, the plot indicates that near-optimal FRIS positioning can be achieved with a small PSO iteration budget and modest swarm.

\begin{figure}[t]
    \centering    \includegraphics[width=0.37\textwidth]{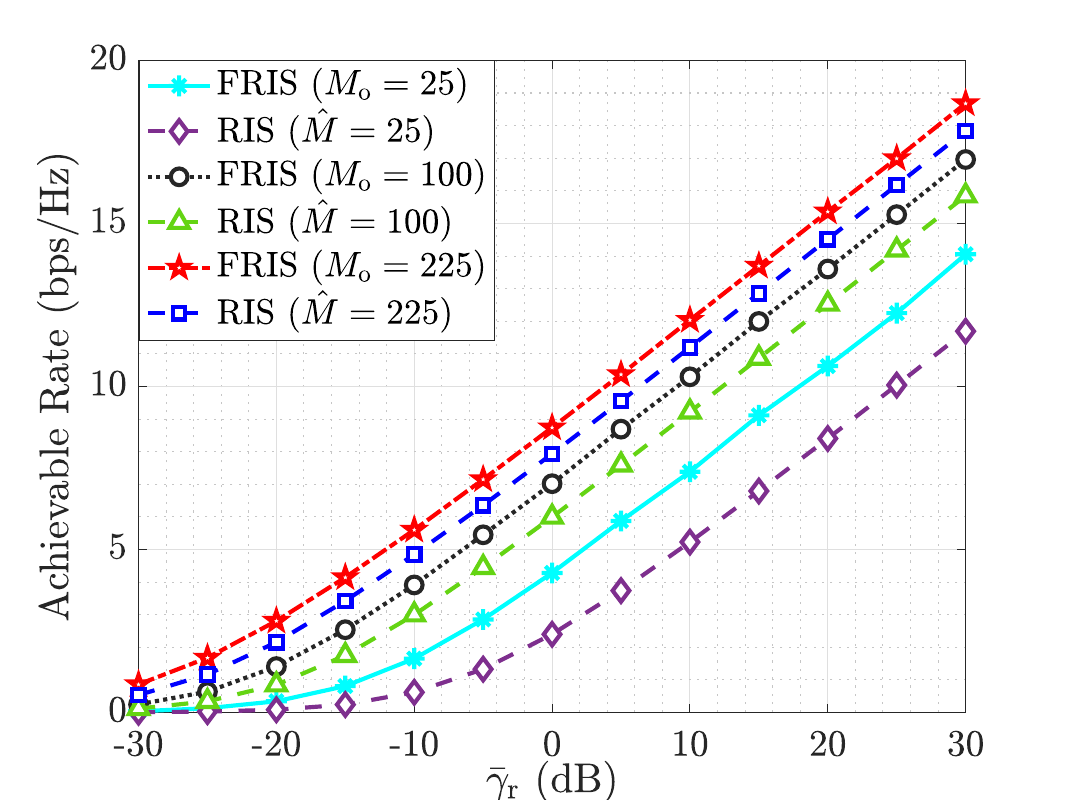}
    \caption{Achievable rate vs. average transmit SNR for different $M_\mathrm{o} $ and $ \hat{M}$.}
    \label{fig:AR_M}
\end{figure}

Fig. \ref{fig:AR_M} reports the achievable backscattering rate versus the average transmit SNR for three array sizes, comparing FRIS with a conventional RIS under the same number of active elements ($M_{\mathrm{o}}\!\in\!\{25,100,225\}$ and $\widehat{M}\!=\!M_{\mathrm{o}}$). Several trends are clear. 
(i) For any SNR, FRIS consistently outperforms RIS, and the gap widens as the active aperture grows: positional reconfiguration lets FRIS place its $m\!\times\!m$ block on stronger spatial clusters, which boosts the coherent sum $\sum_{m\in\mathcal{O}}\!|h_{\mathrm{r},m}||h_{\mathrm{b},m}|$ beyond what a fixed lattice can deliver. 
(ii) Increasing $M_{\mathrm{o}}$ raises the rate for both architectures; more reflecting elements provide larger array gain, yet the returns taper from $100$ to $225$ elements because of spatial correlation and the finite aperture. 

\begin{figure}[t]
    \centering    \includegraphics[width=0.37\textwidth]{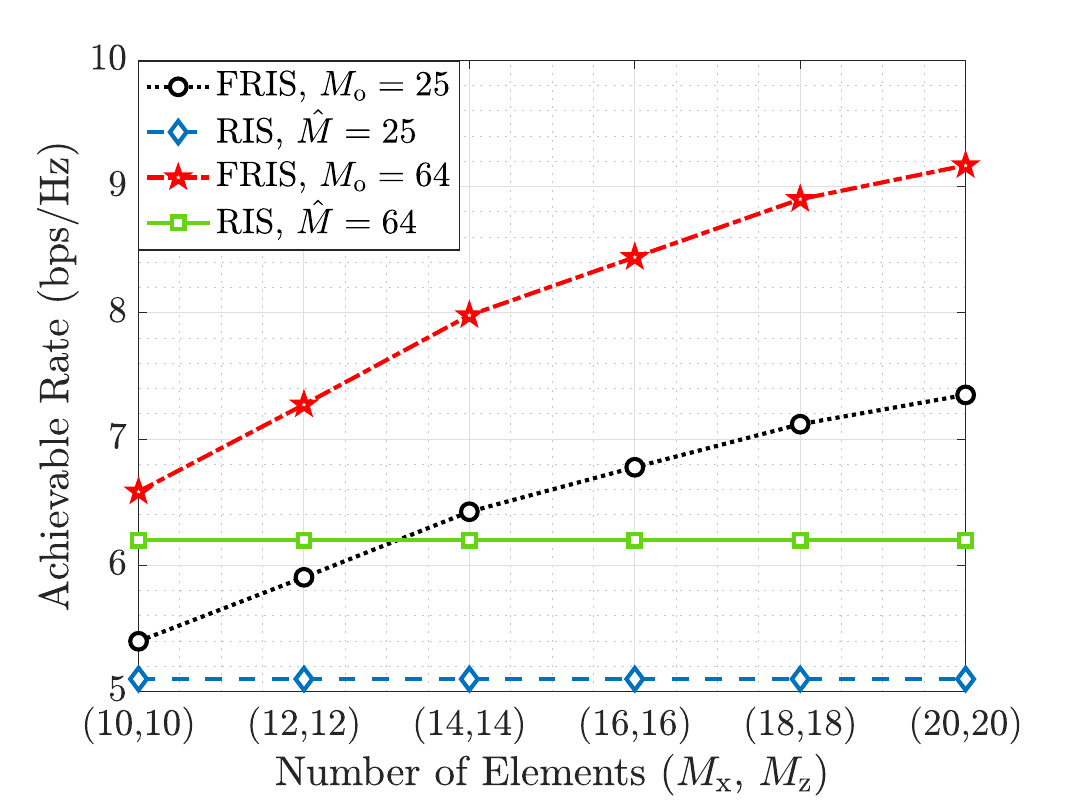}
    \caption{Achievable rate vs. total number of FRIS elements for different $M_\mathrm{o} $ and $ \hat{M}$.}
    \label{fig:AR_Mx_Mz}
\end{figure}

Fig. \ref{fig:AR_Mx_Mz} shows the achievable rate at $\bar{\gamma}_{\mathrm{r}}=10$\,dB as the total FRIS lattice size $(M_{\mathrm{x}},M_{\mathrm{z}})$ grows, for two ON budgets $M_{\mathrm{o}}\!\in\!\{25,64\}$, together with the RIS baselines that use the same number of reflecting elements ($\widehat{M}=M_{\mathrm{o}}$). Two observations stand out. 
(i) \emph{FRIS benefits from a larger host aperture even when the number of active units is fixed.} As $(M_{\mathrm{x}},M_{\mathrm{z}})$ increases, the rate rises monotonically (about $+2$\,bps/Hz for $M_{\mathrm{o}}=25$ and $+2.5$\,bps/Hz for $M_{\mathrm{o}}=64$ from $(10,10)$ to $(20,20)$). The gain comes from positional agility: with more candidate sites across the aperture, PSO can place the $m\!\times\!m$ active block on stronger spatial clusters (where $|h_{\mathrm{r},m}||h_{\mathrm{b},m}|$ is large), improving the coherent sum and thus the rate. The slope gradually tapers, indicating diminishing returns once the “best” regions are already exploitable. 
(ii) \emph{RIS is essentially flat}, because the conventional panel has a fixed $m\!\times\!m$ layout; enlarging the virtual grid does not create additional degrees of freedom. Consequently, the FRIS–RIS gap widens with the total lattice size for a given $M_{\mathrm{o}}$, and is larger for $M_{\mathrm{o}}=64$ than for $M_{\mathrm{o}}=25$, reflecting the compounding effect of more active elements and better placement.

\section{Conclusion } \label{sec_conclusion}
This paper introduced a FRIS–aided AmBC system in which a passive tag communicates with a reader through an FRIS whose positions/activation are optimized while per–element phases are set in closed form. We established a correlated Rician channel model driven by a Jakes' kerne and posed the positional optimization as a non-convex problem that we solved with a lightweight PSO search over the active fluid elements. The resulting framework captures the key benefit of FRIS, positional agility on a fixed aperture, without adding RF chains or breaking AmBC’s low-power ethos.  
Simulations demonstrated consistent and meaningful gains over a conventional RIS baseline across the SNR range and array sizes. 
In particular, FRIS delivers a near-constant rate (or SNR) offset thanks to its ability to place the m$\times$ m ON block on the strongest spatial clusters, and the advantage widens with a larger host aperture even when the number of active elements is fixed. Moreover, the PSO controller converges quickly, typically within a handful of iterations, to a near-optimal layout, making online adaptation feasible for slowly varying channels. 

\section*{Acknowledgment}
{This work has received funding from the SNS JU under the EU’s Horizon Europe research and innovation programme under Grant Agreement No. 101192113 (Ambient-6G) and the EU's Horizon 2022 Research and Innovation Programme under Marie Skłodowska-Curie Grant
No. 101107993.}

\end{document}